\documentclass[preprint2]{aastex}

\shorttitle{PRD models: Explosion}
\shortauthors{Bravo, Garc\'\i a-Senz \& Cabez\'on}

\def\dens{g cm$^{-3}$}

\newcommand{\element}[3]{\ensuremath{^{#2}}#3
}

\begin{document}

\title{Pulsating reverse detonation models of Type Ia supernovae. II: Explosion}
\author{Eduardo Bravo\altaffilmark{1,2}, 
Domingo Garc\'\i a-Senz\altaffilmark{1,2},
Rub\'en M. Cabez\'on\altaffilmark{1},
Inmaculada Dom\'\i nguez\altaffilmark{3}}
\altaffiltext{1}{Dept. F\'\i sica i Enginyeria Nuclear, Univ. Polit\`ecnica de
Catalunya, Diagonal 647, 08028 Barcelona, Spain;   
eduardo.bravo@upc.edu domingo.garcia@upc.edu ruben.cabezon@upc.edu}
\altaffiltext{2}{Institut d'Estudis Espacials de Catalunya, Barcelona, Spain}
\altaffiltext{3}{Depto. F\'\i sica Te\'orica y del Cosmos, Univ. Granada; 
inma@ugr.es}

\begin{abstract}
Observational evidences point to a common explosion mechanism of Type Ia supernovae based on a delayed detonation of a white dwarf.
However, all attempts to find a convincing ignition mechanism based on a delayed detonation in a destabilized, expanding, white dwarf have been elusive so far. One of the possibilities that has been invoked is that an inefficient deflagration leads to pulsation of a Chandrasekhar-mass white dwarf, followed by formation of an accretion shock that confines a carbon-oxygen rich core, while transforming the kinetic energy of the collapsing halo into thermal energy of the core, until an inward moving detonation is formed. This chain of events has been termed Pulsating Reverse Detonation (PRD). In this work we present three dimensional numerical simulations of PRD models from the time of detonation initiation up to homologous expansion. Different models characterized by the amount of mass burned during the deflagration phase, $M_\mathrm{defl}$, give explosions spanning a range of kinetic energies, $K\sim(1.0-1.2)\times10^{51}$~erg, and $^{56}$Ni masses, $M(^{56}\mathrm{Ni})\sim0.6-0.8$~M$_\odot$, which are compatible with what is expected for typical Type Ia supernovae. Spectra and light curves of angle-averaged spherically symmetric versions of the PRD models are discussed. Type Ia supernova spectra pose the most stringent requirements on PRD models.
\end{abstract}


\section{Introduction}\label{intro}

Type Ia supernovae (SNIa) are the most energetic transient phenomena in the 
Universe displaying most of its energy in the optical band of the 
electromagnetic spectrum, where they rival in brightness with their host 
galaxies during several weeks. The  importance of SNIa in astrophysics and 
cosmology is highlighted  by their use as standard (or, better, calibrable) 
candles to measure cosmic distances and related cosmological parameters \citep{per98,sch98,rie01}. 
However, in order to achieve a high precision in the distance determination as 
required, for example, to determine the equation of state of the dark energy 
component of our Universe, it is necessary to understand the physics of SNIa 
explosions. From the theoretical point of view, the accepted model of SNIa 
consists of a carbon-oxygen white dwarf (WD) near the Chandrasekhar mass that accretes matter from a companion in a close binary system. This 
model accounts for the SNIa sample homogeneity, the lack of prominent hydrogen lines in 
their spectra, and its detection in elliptical galaxies. Moreover, massive WDs
 are extremely unstable bodies in which a modest release of energy can produce 
 a huge change of size (i.e. the ratio of binding to gravitational energy is 
 only $\sim15\%$, while in a normal star it is $\sim50\%$). There are two main 
 ingredients of the standard model that are still poorly known: the precise 
 configuration of the stellar binary system and its evolution prior to thermal 
 runaway of the WD \citep{lan00,no03,pie03,han04,bad07,hkn08}, and the explosion mechanism \citep{hil00}. Fortunately, as long as 
 a carbon-oxygen WD reaches the Chandrasekhar mass, its previous evolution is not critical for the explosion because the WD structure is determined by the state of a degenerate 
 gas of fermions that can be described with a single parameter: 
 the mass of the star. This fact leaves the explosion mechanism as the most 
 relevant unknown concerning SNIa.
 
In spite of continued theoretical efforts dedicated to understand the 
mechanism behind SNIa, realistic simulations are still unable to provide a 
satisfactory description of the details of these thermonuclear explosions. Nowadays, there is 
consensus that the initial phases of the explosion involve a subsonic 
thermonuclear flame (deflagration), whose propagation competes with the 
expansion of the WD.  After a while, the corrugation of the flame front  
induced by hydrodynamic instabilities culminates in an acceleration of the 
effective combustion rate. However, the huge difference of lengthscales between the white dwarf ($\sim10^8$~cm) and the flame width ($\lesssim1$~cm) prevents that large-scale numerical simulations can resolve the deflagration front, making it necessary to implement a model of the subsonic flame. The precise value of the effective combustion rate is 
currently under debate, as it depends on details of the flame model. Recent three-dimensional (3D) models calculated by different groups have shown that 
pure deflagrations always give final kinetic energies that fall 
short of $10^{51}$ ergs, while leaving too much unburnt carbon and oxygen close
to the center \citep{rei02b,gam03,gar05,rop07c}. Because both signatures are at odds with observational constraints,  \citet{gam03,gam04,gam05} concluded that the only way to reconcile three-dimensional simulations with observations is to assume that a detonation ignites after a few tenths of a solar mass have been incinerated subsonically (delayed detonation). 

Through a systematic analysis of a large sample of well observed SNIa, \cite{maz07} showed that in spite of the significant dispersion of supernova luminosities, implying different ejected masses of $^{56}$Ni, other characteristic features of SNIa are remarkably homogeneous. In particular, the maximum velocity reached by intermediate mass elements, $\sim11\,000$~km\,s$^{-1}$, indicates a similar extent of thermonuclear burning in all SNIa. Admitting that part of the diversity shown by the SNIa sample
is directly related to the explosion mechanism of Chandrasekhar-mass 
white dwarfs \citep{hblbfg00}, multidimensional 
hydrodynamic simulations of the phenomenon should be able to 
explain the wide range of nickel masses that are seemingly synthesized in different 
events. Lacking multidimensional calculations of the light curve and spectra of these models, 
it is certainly risky to compare directly the outcome of three dimensional explosion simulations with SNIa observations, in order to elucidate the more promising explosion mechanism and discriminate between models. 
However, even acknowledging that a perfect match between the observed luminosity and spectral evolution of SNIa and those deduced from angle-averaged versions of the three dimensional explosion models is not to be expected, there are a few elementary requirements that a model of typical SNIa should pass:
\begin{itemize}
\item There should be no more than a few tenths of unburnt carbon moving 
at low velocities \citep{mhvw03,blh03,koz05}.
\item The upper limit of the ejected mass of $^{56}$Ni should reach $\sim1$~M$_\odot$, after neutronization due to electron captures is properly taken into account \citep{maz07}.
\item Intermediate-mass elements should be generated in amounts exceeding $\sim0.2$~M$_{\odot}$ 
\citep{gar07}.
\item The ejecta should be chemically stratified, as demanded by observations of supernovae and remnants \citep[e.g.][]{bad06,ger07,maz08}.
\item Large clumps of radioactive $^{56}$Ni and other Fe-group elements should not be present at the photosphere 
at the time of maximum brightness \citep{tkbb02}.
\item In order to reproduce the observed width-luminosity relationship the kinetic energy and the total mass burned have to be in the ranges $K\sim\left(1.0-1.4\right)\times10^{51}$~ergs and $M_\mathrm{burnt}\sim1.0-1.2$~M$_\odot$, respectively \citep{woo07}\footnote{\citet{woo07} also required that the inner layers of ejecta $(\lesssim0.8$~M$_\odot)$ should be fairly well mixed, but such extent of mixing can be in conflict with what is required to explain SNIa spectra both in the optical and in the infrared \citep{mot06,ger07,maz08}}. 
\end{itemize}

\citet{bar08} calculated detailed non-LTE synthetic spectra of angle-averaged spherically symmetric versions of Pulsating Reverse Detonation (PRD) models. As compared with early, maximum, and post-maximum spectra of well observed SNIa, the PRD models produced an excessively prominent C~II feature, and the spectra appeared too red due to the presence of iron-peak elements in the outer layers. \citet{bar08} pointed out that fitting of real SNIa spectra might improve if some fraction of the iron-peak elements moving at high velocities were substituted by intermediate-mass elements. 

In a companion paper (Bravo \& Garc\'\i a-Senz, 2009; hereafter paper~I) we have analyzed the conditions for detonation ignition after a pulsating phase of a white dwarf. A confined detonation ignition (CDI) can be induced by an accretion shock, formed around a carbon-oxygen rich core after pulsation following subsonic burning of a mass  $M_\mathrm{defl}<0.3$~M$_\odot$. The mechanism of CDI is robust, and detonation ignition conditions are reached non-marginally within the shocked flow for a wide range of initial configurations. However, if the nuclear energy released before the pulsation is close to the white dwarf binding energy, $M_\mathrm{defl}\sim0.29$~M$_\odot$, the accretion shock is not efficient enough to initiate a detonation in a mixture of carbon and oxygen. In that paper we have also shown that if C-O matter is contaminated by a small amount of He ($X(\mathrm{He})\sim0.05-0.10$) the conditions for a successful CDI are relaxed substantially, making even easier the detonation of the core. A small cap made of He might rest atop of the white dwarf at the moment of thermal runaway as a result of previous accretion from the companion star in the binary system. This helium could either have been accreted directly from a He star or result from the nuclear processing of accreted hydrogen. \citet{hk96} suggested that the presence of a $\sim0.01$~M$_\odot$ cap of He would improve the match of calculated and observed light curves. There have been some claims of detection of He~I lines at 2.04~$\mu$m and 1.052~$\mu$m in the spectra of the spectroscopically normal SN 1994D and other SNIa \citep{cu94a,cu94b,no03,pi08}, although its identification is very uncertain \citep{wh98,ml98,ha99}. In a few test calculations shown in the present paper we have assumed the presence of such a small cap of He.

In the present work we give details of the detonation phase of PRD models computed in three dimensions, together with an analysis of the nucleosynthesis and the bolometric light curve obtained from angle-averaged spherically symmetric versions of such models. We also discuss the consequences that departures from spherical symmetry of the ejecta can have over the spectra. In Section~2 we explain the methodology used to compute the three-dimensional supernova models presented in this paper. In Section~3 the evolution of the white dwarf during the detonation phase is described, while in Section~4 the properties of the homologous expanding ejecta are analyzed. Finally, our conclusions are given in Section~5.

\section{Methods and models}

The simulations were performed with a three-dimensional Lagrangian 
Smoothed Particle Hydrodynamics (SPH) code suited to the modelling of Type Ia 
supernovae \citep{gar98,gar99}. Contrary to Eulerian methods in which usually the spatial resolution is a free numerical parameter and the mass resolution is determined by the local density, in SPH it is the mass resolution of the calculation which is fixed\footnote{Although it is possible to dynamically increase the mass resolution during the SPH calculation using a technique known as {\it particle splitting}, that is the equivalent to AMR methods in Eulerian codes, we have not used it in the calculations reported in the present paper.}. In SPH, the number of interpolating neighbors is what determines the spatial resolution of the calculation. In our models, the number of neighbors was variable in the range 40-90, while the total number of equal-mass particles was $N = 250,000$. The maximum spatial resolution reached during the simulations of the detonation phase of the PRD models was $9-12$~km, depending on the model (the radius of the star at that epoch was $\sim40,000 - 95,000$~km). Further details of the numerical techniques applied in the SPH code are provided in the appendix.

The physical ingredients included in the hydrocode allow to describe accurately all the phenomena relevant to thermonuclear supernovae.
Gravity was calculated by means of the hierarchical tree method \citep{her87}, retaining up to the quadrupole terms in the 
multipolar expansion. In order to improve the accuracy of the calculation, the gravitational interaction between close particles was computed directly.
The equation of state includes contributions of partially degenerate and 
relativistic electrons with pair corrections, ions considered as an ideal gas 
plus Coulomb corrections and  radiation. The nuclear binding energy and 
electron capture rates \citep{ffn82,mar00} on matter in nuclear statistical equilibrium 
(NSE) were interpolated from a table that uses density, temperature and 
electron mole number as input. Weak interactions in general, and electron 
captures in particular, are necessary for an accurate description of 
thermonuclear supernovae because they affect the dynamical evolution by decreasing the total number of electrons that support the white dwarf structure. The mean 
electron mole number of the ashes, $\Delta Y_\mathrm{e}^\mathrm{a}$, determined by electron captures during the 
deflagration phase is $\Delta Y_\mathrm{e}^\mathrm{a}\sim0.484$~mol~g$^{-1}$. This neutronization translates into 
a decrease of electron pressure of as much as $\Delta p_\mathrm{e}^\mathrm{a}/p \sim (8/3) \Delta Y_\mathrm{e}^\mathrm{a} \sim 4\%$, of the same order as the
increase in pressure due to incineration of carbon and oxygen at a density of 
$10^9$~\dens, $\Delta p_\mathrm{burn}/p\sim10\%$.

The nuclear network consumed a huge part of the computational 
resources chiefly due to the very small time-steps required to integrate 
accurately the nuclear kinetic equations, and because at some point of the 
calculation there were tens of thousands of particles undergoing simultaneously 
nuclear combustion. A small network of 9 isotopes: n, $^{1}$H, $^{4}$He, 
$^{12}$C, $^{16}$O, $^{20}$Ne, $^{24}$Mg, $^{28}$Si, and $^{56}$Ni \citep{woo86} allows 
to calculate the approximate nuclear energy input and a rough nucleosynthesis 
with a moderate load of CPU time. The performance of this network was 
extensively tested in \citet{tim00}. An operator-split approach was used to make the 
calculation feasible. As soon as the characteristic nuclear time-step of a 
given particle became smaller than a prescribed fraction of the dynamical time-step its chemical 
evolution was followed isochorically, and both the nuclear network and the equation of 
state were decoupled from the hydrodynamic evolution until the dynamical 
time-step was recovered. 
When the temperature of a mass element reached $5.5\times10^9$~K, NSE was assumed. 
Once attained, NSE was maintained as long as the temperature stayed above $2\times10^9$~K, providing a nuclear energy generation rate accurate enough for the hydrodynamical simulations. 
Weak interactions during NSE determine the electron mole number, $Y_\mathrm{e}$, 
whose evolution was calculated at each timestep by solving the 
equation: $\mathrm{d} Y_\mathrm{e}/\mathrm{d} t = \Sigma_i \lambda_i Y_i$, 
where $\lambda_i$ stands for all kind of weak interactions:
$\lambda_i > 0$ for $\beta^-$ disintegrations and e$^+$ captures, and 
$\lambda_i < 0$ for $\beta^+$ disintegrations and e$^-$ captures, while the 
molar fractions, $Y_i$, were set by the NSE equations.

\subsection{Testing the code}\label{testcode}

We have tested the performance of our SPH hydrocode simulating two three-dimensional delayed detonation scenarios available in the scientific literature, which had been computed using quite different numerical methods (both for resolving the hydrodynamical equations and to describe the nuclear flames). We have chosen delayed detonation models that bear many similarities with the PRD scenario, i.e. those in which a delayed detonation is formed at a low density and propagates inwards through a carbon-oxygen rich white dwarf core. Our reference models are the Gravitationally Confined Detonation (GCD) model Y100 from \citet{pl07}, and the turbulent delayed detonation model DD\_005 from \citep{rop07}. The first one begins with the incineration of a small off-centered bubble which floats to low-density regions consuming a few hundredths of a solar mass to finally break-up the white dwarf surface. The combustion ashes slide through the surface to finally concur at the opposite pole igniting a detonation which processes most of the white dwarf to NSE, mainly in the form of $^{56}$Ni. The turbulent delayed detonation model DD\_005 is based on a physical criteria of detonation ignition, namely the transition from a laminar flame to the distributed regime of nuclear combustion. We have computed models GCD1 and TURB7 using similar initial models and physical assumptions as for Y100\footnote{Note that Y100 is a two dimensional model, while GCD1 was computed in 3D} and DD\_005, respectively. Model GCD1 started from a single incinerated bubble of radius 53~km located 100~km off-center. Model TURB7 started from the same initial model as model DF29 in paper~I, but it incorporated a prescription for turbulent acceleration of the burning rate according to the local velocity fluctuations. Although the initial configuration of TURB7 is not the same as in DD\_005, both models burnt similar masses subsonically, that were distributed in a small number of ash clumps, and used the same criteria for the deflagration-to-detonation transition. As we want to test the ability of the code to deal with detonation propagation, the similarity of the global properties of the white dwarf at the end of the deflagration era is what makes meaningful the comparison of models DD\_005 and TURB7.

The outcome of the additional models is summarized in Table~\ref{tab1}, where we give the amount of mass burned during the initial deflagration phase (when available), the final kinetic energy, and the ejected mass of $^{56}$Ni, NSE elements, carbon plus oxygen, and intermediate-mass elements (details of these simulations and the numerical setup used will be presented in a forthcoming paper,  Garc\'\i a-Senz et. al. 2009). The mass of $^{56}$Ni in the reference models is just an upper limit because electron captures were not included in the hydrodynamic calculations. As can be seen in the Table, the matching between our models and the reference ones is excellent, especially when one compares GCD1 with Y100, which are nearly twin models. Model TURB7 reproduces quite well the main properties of model DD\_005, but some small differences remain in the synthesized masses of intermediate-mass elements (IME) and NSE elements. The reason for these discrepancies may reside in different approaches to the description of the flame as well as on the effect of the electron captures on the dynamics of the explosion. Even though neither DD\_005 nor Y100 included electron captures, the amount of mass burned at high densities during the deflagration phase was much larger in the turbulent delayed detonation model (0.38~M$_\odot$ vs $\sim0.06$~M$_\odot$), hence the impact of the neutronization on the explosion dynamics should be also larger in TURB7. The final mean electron mole number in GCD1 and TURB7 was $Y_\mathrm{e}= 0.4964$ and $Y_\mathrm{e}=0.4932$, respectively, i.e. the neutronization in the turbulent model was about double than in the other one.

\subsection{Pulsating reverse detonation simulations}

Our working hypotheses are that during the deflagration phase it is burnt an insufficient amount of mass to unbind the white dwarf, and that the outcome of the explosion is predominantly determined by $M_\mathrm{defl}$. The simulations of the deflagration phase started from initial models consisting on spherically symmetric cold ($T = 10^8$~K) 
isothermal white dwarfs built through a fourth order Runge-Kutta integration. 
The one-dimensional profile was mapped to a three-dimensional distribution of $N$ particles and 
afterwards relaxed with the procedure explained in \citet{gar98}. Once a stable enough 
configuration was obtained, the temperature of a small subset of particles was 
artificially increased up to the self-consistent value of NSE temperature at 
such density (assuming isochoric adiabatic burning) and the ensuing dynamical 
evolution was followed by means of the SPH hydrocode.

The main properties of our PRD models are given in Table~\ref{tab2}. For clarity, we have changed the name of the models with respect to previous publications. In the present work, PRD models are designated with the acronym PRDnn, where 'nn' gives the hundredths of a solar mass burned during the initial deflagration phase. For instance, in model PRD18 the mass burned subsonically before the pulsation was $M_\mathrm{defl} = 0.18$~M$_\odot$ (this model was called PRD6 in \citealt{bra06} and \citealt{bar08}, while the present model PRD14 was called PRD5.5 in \citealt{bar08}). 

\begin{figure}
\plotone{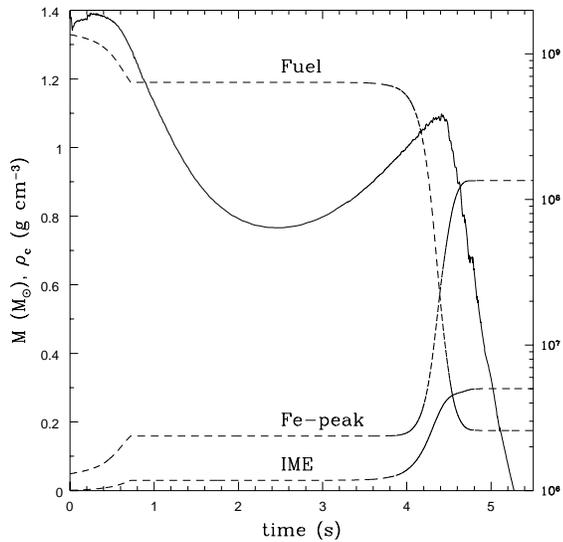}
\caption{Evolution of model PRD14 through the deflagration, pulsation, and detonation phases. Shown are the evolution of the total masses of fuel, intermediate-mass, and Fe-peak elements (left axis scale), and the central density (right axis scale)
}\label{newfig1}
\end{figure}

\begin{figure}
\plotone{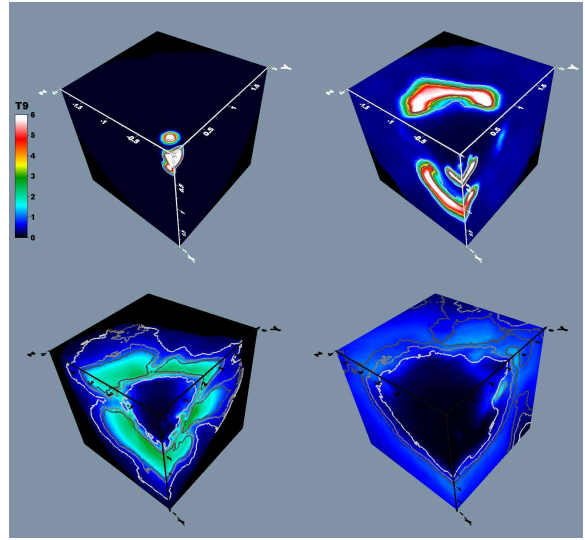}
\caption{Evolution of temperature (color map, in $10^9$~K) and fuel mass fraction (contours) in an octant of model PRD14 
during the deflagration and pulsation phases. The C+O mass fraction contours shown are
$X=0.3$ (black), 0.5 (dark gray), and 0.8 (light gray). Axes units are thousands of km. Times shown are (from left to right and top to bottom): $t=0$, 0.5, 1.0, and 2.0~s. Note that the size of the axes changes: $2\,000$~km in the top row images, and $6\,000$~km in the bottom row. The white dwarf radii at the times shown are: $1\,900$, $2\,300$, $6\,500$, and $16\,600$~km
}\label{newfig3}
\end{figure}

The evolution of the central density and the chemical composition of model PRD14 is shown in Fig.~\ref{newfig1}. At the beginning combustion propagated subsonically as a deflagration wave, with an effective combustion rate low enough to ensure that the flame quenched before the white dwarf became unbind. During the first seconds of the ensuing pulsation the incinerated 
bubbles floated to the surface, leading to composition inversion, i.e. the internal volume, which we call hereafter the core,  was plenty of cold fuel, i.e. carbon and 
oxygen, while the ashes of the initial combustion, 
mostly hot iron and nickel, were scattered around (see Fig.~\ref{newfig3}). Shortly after, the expansion of the core came to an end while part of the envelope re-collapsed. Eventually, an accretion shock was born because of the impact of the in-falling material onto the carbon-oxygen core. Upon re-collapse nuclear reactions light again a few tenths of a second before the time of maximum compression ($\sim4.4$~s). In this paper, our interest is focused on the events that ensue after a detonation is ignited due to the energy transferred to the core through the accretion shock \citep[the deflagration phase has been thoroughly described in][]{gar05}. 

\section{Description of the detonation phase of PRD models}

In paper~I we have shown that after a pulsation the white dwarf reaches conditions favorable for a CDI if $M_\mathrm{defl}$ is well below $\sim0.3$~M$_\odot$ and nuclear reactions are turned off. However, when nuclear reactions are included in the calculation the release of nuclear energy contributes not only to temperature increase but also to raising the gas pressure, thus disturbing the dynamical evolution during the collapsing phase of the pulsation. In the present simulations we have allowed the detonation to ignite spontaneously at the time and place determined by the SPH code as a result of the hydrodynamical evolution coupled to the nuclear network. This approach must be looked with caution as we cannot resolve the detonation front thickness, hence our simulations might not describe accurately the process of detonation formation. However, as we will see later in Sect.~\ref{ejecta}, the explosion properties are quite independent of the details of the detonation initiation, hence our approach is justified.

\subsection{Initial steps of detonation propagation}

It is interesting to analyze the initial steps of detonation propagation, keeping in mind the cautionary remark given above about the capabilities of our code to describe accurately the birth of a detonation front. Figure~\ref{newfig4} shows the evolution of a zoomed slice across model PRD14 for two times: just after the nuclear timescale becomes shorter than the hydrodynamical timescale, and 0.1~s later (top images). The maximum temperatures achieved in the region are high enough ($T\gtrsim5\times10^9$~K) for explosive silicon burning, whereas carbon and oxygen are destroyed in a shorter timescale once the temperature rises above $\sim3\times10^9$~K as highlighted by the fuel contours. The thermal gradient in the radial direction is large, on the order of $100-200$~K~cm$^{-1}$, although it is influenced by the resolution of the numerical calculation. Shock fronts propagate ahead and close to the thermal waves, as revealed by the sculpting of the isobars following the contour of the high temperature regions and by the large pressure gradient in front of the thermal front (bottom left image): the detonation is born. 

The performance of the detonation capture algorithm is illustrated in the bottom right image on Fig.~\ref{newfig4}, where it is shown the region where the adaptive elliptical interpolator gives a sharper detonation front. As can be seen, a detonation is detected all around the thermal wave except the outermost part characterized by a shallow thermal gradient. Figure~\ref{newfig5} displays a complementary view showing the location (in fact, the projection onto the $yz$ plane) of the actual SPH particles within the detonation front, i.e. those for which the detonation capture condition has been verified. The color map in the same Fig. shows the smoothing length, whose value at the detonation front is of order $30$~km. 

\subsection{Burning of the white dwarf core}

In the PRD scenario, the last phase of the explosion starts when the converging reverse detonation 
wave is launched. 
Once initiated, the detonation is self-sustained by the nuclear energy release from burning of the 
carbon-oxygen fuel that pervades the hydrostatic core. 

The evolution of the radii of spherical shells of given Lagrangian mass is shown in Fig.~\ref{newfig2}, together with the evolution of the detonation front. Because the model does not have spherical symmetry, the location of the detonation front is only approximate. In that figure, the inner and outer boundaries of the detonated region have been defined as the radius of the shells for which at least half of the carbon and oxygen remaining after the deflagration phase had been processed by the detonation wave (see Fig.~\ref{figtem} and the next paragraph for a complementary view of the evolution of the detonation wave). It can be seen that the inwards propagation of the thermonuclear burning is very fast, as corresponds to a detonation front. This detonation front processes most of the inner core of the white dwarf, but it stalls before arriving to the very central regions and fails to burn the innermost few hundredths of a solar mass.

Fig.~\ref{figtem} illustrates the evolution of the white dwarf core during the propagation of the reverse detonation for model PRD14, with a view complementary to the one in Fig.~\ref{newfig2}. At the time of the first snapshot, $t=3.8$~s, the accretion shock is about to form. The C+O mass fraction contours maintain a high degree of spherical symmetry in the core, increasing their value steeply towards the center (lighter contours belong to larger C+O mass fractions). In the first snapshot, a hot spot is visible in the $yz$ plane at $r\sim2\,000$~km, which is a result of a slight asymmetry in the collapsing flow (the temperature and chemical composition maps a few tenths of a second before do not show any sign indicating the imminent formation of a hot spot). The combination of high temperatures and abundant fuel culminated in the formation of a detonation already visible in the second snapshot. At the same time, the accretion shock had heated the fuel in a hot spherical shell located at $r\sim3\,000$~km up to $T\gtrsim1.5\times10^9$~K. The lateral spreading of burning was not due to detonation propagation, but it was caused by the accretion shock, that continuously fed mechanical energy into the core sparking detonations all around the hot shell. 
During the first 0.3~s after detonation ignition the 
accretion shock remained stationary, close to the core. Afterwards, the 
overpressure generated by the nuclear energy released pushed out the accretion 
shock, that detached from the dense core (last snapshot). In 
the outer layers of the core the density was low enough to allow for 
incomplete silicon burning and leave a composition rich in intermediate mass elements, mainly Si, S, Ca and Ar, while in the inner regions the burning proceeded all the way up to $^{56}$Ni (the central density at the moment of 
formation of the detonation was in the range $(0.5-4)\times10^8$~\dens). 

In our simulations the detonation propagated inwards burning all the fuel except for a small volume close to the center of the white dwarf. The unburnt mass, $\sim0.08$~M$_{\sun}$, and the central density, $\rho_\mathrm{c}\sim4\times10^8$~\dens, 
give a volume of $\sim4\times10^8$~km$^3$ or a typical lengthscale of 
the unburned spot of $\sim450$~km. Our resolution at the center at that time was of 
order 11~km, implying that the detonation failure at the center of the white dwarf was not due to insufficient numerical resolution. The physical reason for detonation failure at the center is the expansion
of the external, already detonated, core matter. As the detonation wave is
accessible to the shocked matter through sound waves, the rarefaction wave originated by expansion of the core caught and weakened the shock and finally 
quenched the burning close to the center. It took a few
tenths of a second since detonation initiation to release enough nuclear energy
to overcome the impact pressure of accreting matter and allow the detonated
matter to expand appreciably. As a result of such delay most of the core matter was 
detonated, leaving only a small mass of unburned fuel near the center. The amount of unburned carbon and oxygen is sensitive to the density structure at the moment of detonation formation, being the largest for the less energetic model PRD26 (see Table~\ref{tab2}).

\subsection{Sensitivity and resolution}

We have checked the sensitivity of the results of our simulations to the numerical method, in particular to the implementation of artificial viscosity, and to the spatial resolution. 
As it is well known, artificial viscosity introduces numerical dissipation in the evolution equations, which can generate enough entropy at shocks to induce an artificial detonation. 
We have computed two additional models starting from $M_\mathrm{defl}=0.18$~M$_\odot$, 
one with the artificial viscosity parameters enlarged by a factor two, the 
other with the parameters divided by the same factor. In both cases, the code 
included an algorithm to inhibit combustion inside the artificially widened shock fronts. The detonation 
developed in both cases in the same way as in our standard calculation,
irrespectively of the values taken by the artificial viscosity parameters. The final kinetic energies and $^{56}$Ni masses synthesized differred by less than 8\% with respect to our standard model PRD18.

We also checked the sensitivity of the results to the numerical resolution by running model PRD18 with the same number of particles but reducing the number of neighbors, $n_0$. The smaller number of neighbors translated into a smaller smoothing length, and thus in an increased capability of the code to 
represent small scale features. Such procedure allowed us to reach a maximum resolution of 
4~km during detonation propagation. Once more, the detonation developed in the same way as in our standard 
calculation. As a drawback, the small number of neighbors originated numerical instabilities in the outermost low-density regions of the ejecta. These instabilities prevented the model to reach the homologous expansion phase, making it impossible to compare the final outcome with that of our standard PRD18 model.

Finally, we checked if the outcome of the explosion was sensitive to the presence of a small mass of He atop of the white dwarf. This helium mixed with the underlying layers of carbon and oxygen during the pulsation and was present at the detonated layer at ignition time. 
By adding either a total He mass of 0.005~M$_{\sun}$ (i.e. a 0.36\%) or 
0.010~M$_{\sun}$ (i.e. a 0.72\%) to model PRD18 the kinetic energy and the mass of $^{56}$Ni synthesized in the explosion
changed negligibly ($\lesssim4\%$) with respect to the standard calculation of the same model. Thus, we conclude that the eventual presence of He does not introduce additional uncertainties in the outcome of the PRD models.

\section{Nucleosynthesis, light curves and spectra: asymmetries, inhomogeneities}\label{ejecta}

In Table~\ref{tab2} there are given the properties of the PRD models we have computed, covering the range of subsonically burned masses $M_\mathrm{defl}=0.14-0.26$~M$_\odot$. The kinetic energy at infinity spans the range $\left(1.0-1.2\right)\times10^{51}$~erg, in good  agreement with requirements deduced from the SNIa light curve shape (see Sect.~\ref{intro}). In principle, it should be possible to obtain PRD models with still smaller values of $M_\mathrm{defl}$, down to $\sim0.10$~M$_\odot$ or even less, which according to the trend displayed in Table~\ref{tab2} would translate into larger kinetic energies and $^{56}$Ni masses. However, as the accretion shock in model PRD14 already forms quite close to the total mass of the white dwarf, it is to be expected that the outcome of the explosion for smaller $M_\mathrm{defl}$ will not be much different from that of model PRD14. On the other side, values of $M_\mathrm{defl}$ slightly larger than those in model PRD26 would probably give rise to explosions with slightly smaller kinetic energies, substantially smaller $^{56}$Ni masses and much larger intermediate-mass elements yields. Unfortunately, due to the failure to maintain a steady detonation in model PRD29 we have not been able to verify this last point.

As seen in the last snapshot of Fig.~\ref{figtem}, the detonation smooths the thermal and chemical structure of the white dwarf, wiping out the imprints left by the deflagration phase in the central volume of the ejecta. The detonation also endows the mechanical structure of the ejecta with a large degree of spherical symmetry. 
Fig.~\ref{figdens} shows the homologous density profile of model PRD14 at a time of 34~s since the initial thermal runaway. The solid line represents the angle-averaged density obtained in 1\,000 shells evenly distributed in mass, while the error bars give the standard deviation of the particles density. The small dispersion is an indication of the high degree of 
spherical symmetry. The 
high-velocity low-density external material is best approximated by an 
exponential density law, whereas the central region can be approximately described as 
a nearly uniform density core whose mass is $\sim0.82$~M$_\odot$. The exponential tail begins at a 
velocity $\sim10\,000$~km~s$^{-1}$, where the intermediate-mass elements (Si, S, 
Ca, ...) are abundant: $X_\mathrm{IME}\gtrsim0.2$. The innermost 0.08~M$_{\sun}$, composed mainly by unburned 
C-O, is denser than the uniform core by up to an order of magnitude.

A comparison of the density profiles of the last computed models (normalized by $t^3$) is presented in
Fig.~\ref{figdens2}. 
The larger kinetic energy of model PRD14 with respect to PRD18 produces a more extended tail in the former. The beginning of the exponential tail matches quite approximately the location where the detonation was ignited in each model (see values of $v_\mathrm{deto}$ in Table~\ref{tab2}). Whereas the transition from the central uniform density region to the exponential tail is quite smooth in model PRD14, it becomes increasingly abrupt as $M_\mathrm{deto}$ and $v_\mathrm{deto}$ decrease. In model PRD26 there is not a uniform density core, which is substituted by a density inversion below $M_\mathrm{deto}$. In models PRD18 and PRD26 the density of the innermost region rich in fuel is $\sim4$ times larger than the density of the nearly homogeneous core, a factor much lower than that found in model PRD14.

\subsection{Nucleosynthesis}

Figs.~\ref{fig5} and \ref{fig05} depict three-dimensional views of the chemical structure of the ejecta. In Fig.~\ref{fig5} a rendering of the final distribution of $^{56}$Ni in model PRD14 is shown. The central volume is homogeneously filled with $^{56}$Ni, except for a small ball made of carbon and oxygen that is not visible in this image, surrounded by intermediate-mass elements and unburned carbon and oxygen. Small and medium size clumps of $^{56}$Ni and other Fe-group elements pervade the outer layers of the ejecta. These clumps have a double origin: some of them were produced during the initial deflagration phase and migrated to the external layers of the white dwarf during the expanding phase of the pulsation, while others detached from the incinerated core after the end of the detonation phase. These clumps are the end product of the impact of the detonated core into the surrounding low-velocity expanding layers. Thus the final distribution of $^{56}$Ni comes from two different regimes: a smooth inner region, rich in $^{56}$Ni and other Fe-group elements, and a clumpy outer region where radioactive and stable Fe-group and intermediate-mass elements share the volume with carbon and oxygen. A complementary picture of the chemical structure is given in Fig.~\ref{fig05}, in which the mean molar weight in an octant of model PRD14 is displayed. It shows the same gross features seen in Fig.~\ref{fig5}, but here more details are visible, as for instance the central ball made of carbon and oxygen. Nevertheless, carbon and oxygen are predominantly located at the outermost layers of the ejecta.

In order to compute even coarse light curves and spectra of SNIa models and to perform a preliminary comparison with observations, it is necessary to obtain the isotopic yields up to $A\sim60$. As our SPH code implements just a 9-isotopes nuclear network, we have post-processed the thermodynamic history of the particles to derive the nucleosynthetic output of each explosion model. 
For each computed model, the SPH particles were classified into three groups: 1) those whose temperatures never exceeded $10^9$~K, that were assumed to keep their initial composition unchanged, 2) those that reached temperatures in excess of $T_\mathrm{NSE}=5.5\times10^9$~K, that were assumed to attain NSE composition and, 3) the remaining particles, that were candidates to experience incomplete explosive burning. The nucleosynthesis of each particle belonging to the last group was computed by integration of the nuclear kinetic equations along the thermodynamic trajectory recorded for that particle during the SPH simulations. The method followed for the integration of the nuclear equations was described in \citet{cab04}. Finally, the particles belonging to the NSE group were again divided in two categories according to the density at which they left NSE. Those particles that cooled below $2\times10^9$~K at densities in excess of $10^8$~\dens were assigned the chemical composition in NSE at these temperature and density, and at the electron mole number given by the SPH model. Conversely, particles that left NSE at densities below $10^8$~\dens were assumed to experience an alpha-rich freeze-out, and their final chemical composition was obtained by integrating the nuclear network following the cooling path of the particle, starting from $T=T_\mathrm{NSE}$ and the chemical composition of NSE matter at that temperature. 

The mass of $^{56}$Ni synthesized in each PRD model can be found in Table~\ref{tab2}, while Table~\ref{tab3} gives the final abundances after radioactive decays, where we have summed up the abundances of the isotopes of each element and we show only the most abundant species in the ejecta. The amount of $^{56}$Ni produced is in the range $0.6-0.8$~M$_\odot$, which is enough to explain the light curves of typical SNIa. As pointed before, models with $M_\mathrm{defl}$ in excess of $\sim0.27$~M$_\odot$ might produce lower amounts of $^{56}$Ni that might account for sub-luminous SNIa, but we have not been able to verify it in the present work. The total mass of IME synthesized in the three computed models spans a similarly narrow range, $M(\mathrm{IME})\sim0.20-0.26$~M$_\odot$. 

Figures~\ref{figcomp1}-\ref{figcomp3} show the final (after radioactive decays) chemical profiles of the three PRD models as a function of the ejecta velocity. The location of the accretion shock, $v_\mathrm{acc}$, and the CDI, $v_\mathrm{deto}$, left no imprint on the chemical profile. The dispersion of the mass fractions agrees with the qualitative view obtained with Figs.~\ref{fig5} and \ref{fig05}: the inner volume is chemically stratified while the outer layers display large scatters. The transition between both regimes takes place at an ejecta velocity of $v\simeq10\,000$~km~s$^{-1}$ in model PRD14 and $v\simeq8\,000$~km~s$^{-1}$ in model PRD18, while in model PRD26 the scatter is already high at a few $1\,000$~km~s$^{-1}$. The abundance of intermediate-mass elements increases towards the surface of the ejecta in all models. In model PRD14 the IME amount to a significant mass fraction at velocities in excess of $v\simeq10\,000$~km~s$^{-1}$, and are dominant above $v\simeq20\,000$~km~s$^{-1}$. The large mass fraction of Fe-group elements at high velocities is a concern because its expected spectral signatures are not seen in SNIa. The same is true for models PRD18 and PRD26. In model PRD18 the abundance of carbon plus oxygen increases monotonically up to the surface where it amounts to $X(\mathrm{C+O})\sim0.30$. The C+O profile in model PRD26 shows a peak at the location of the CDI, $v_\mathrm{deto}=8\,000$~km~s$^{-1}$, and declines at larger velocities. Although this model has the largest mass of IME, Fe-group elements are still dominant through the outermost layers.

In Fig.~\ref{fig12} it is shown the detailed chemical composition of model PRD14 after radioactive decays as a function of the velocity. 
We stress that the poor chemical stratification of the ejecta is mainly due to the mixing between the detonated core material and the surrounding matter that is produced during the initial phases of expansion following the end of the detonation era. Such a mixing is expected to happen in any model in which a detonation propagates inwards from a point close to the surface of the star, as is the case in the two test models presented in Sect.~\ref{testcode}. This is illustrated in Fig.~\ref{vturb}, that shows the chemical profiles of model TURB7 at two times: just after the end of detonation and when the ejecta is in the homologous expansion phase. It is plain to see that the chemical species are redistributed through the outer $\sim0.6$~M$_\odot$ between both times, with the result that Fe-group elements replace C-O and IME in the outer layers. Thus, the final chemical profile of TURB7 displays also a large mass of Fe-Ni moving at high velocities.

\subsection{Light curves}

Multidimensional thermonuclear supernova models assuming different explosion mechanisms have been shown to reproduce reasonably well the gross features of broadband light curves of normal Type Ia supernovae \citep{bli06,kp07,rop07c}. The model light curves in these works were obtained from one and two-dimensional averaged versions of two and three-dimensional supernova models. An alternative approach is to compute {\sl bolometric} light curves from spherically symmetric angle-averaged versions of three-dimensional SNIa models. Bolometric light curves should be less sensitive to the averaging procedure than broadband light curves, as the first ones are not affected by shifting of photon wavelengths in and out of monochromatic bands, as due for instance to iron line blocking.

Bolometric light curves have been computed from angle-averaged versions of the PRD models by means of the one-dimensional code described in \citet{bra93,bra96}. In general, the bolometric light curves obtained with this code during the pre-maximum and maximum phases are in fairly good agreement with those computed by directly solving the radiative transfer equations \citep[see][]{hoe93}. 
Although our PRD models reach the peak of luminosity slightly earlier than expected ($\sim12$~days), the maximum luminosity, $L_\mathrm{max}$, and $^{56}$Ni mass,  $M \left(^{56}\mathrm{Ni}\right)$, match almost perfectly the empirical relationship derived by \citet[their Eq.~6]{str05}. In Fig.~\ref{figm15m56} we show a plot of $M (^{56}\mathrm{Ni})$ vs the bolometric light curve decline parameter $\Delta m_{15}$, both from the present three-dimensional models as well as from observational estimates for a set of well observed SNIa \citep{str06}. Models PRD14 and PRD26 reproduce nicely the observational trend, while model PRD18 declines too fast in comparison to SNIa having a similar mass of $^{56}$Ni. Models GCD1 and TURB7 synthesize a large amount of $^{56}$Ni and, due to their large kinetic energy and the presence of radioactive nuclei up to the surface of the ejecta, decline much faster than observations seem to indicate (the main parameter determining the width of the bolometric light curve is the photon diffusion time, $t_\mathrm{d}$, that is in turn affected by the kinetic energy: $t_\mathrm{d}\propto K^{-1/4}$, \citealt{woo07}). 

\subsection{Spectra}

\citet{bar08} obtained detailed spectra of angle-averaged one-dimensional versions of several PRD models, from slightly before up to a few days after maximum light. They concluded that the synthetic spectra of the PRD models did not match the spectra of a few typical well-observed SNIa. The main disagreements found by \citet{bar08} were due to the presence of Fe-group elements in the outer layers of the ejecta, which translated in an excessive flux at long wavelengths in the computed post-maximum spectra. There was also an excess of carbon, that resulted in a prominent C~II 6580{\AA} line at 13 days that was too red by $\sim250${\AA} and lacked the correct shape\footnote{We note that the C~II line wavelength disagreement might be reduced by $\sim100${\AA} if the epoch of the computed spectra were a few days earlier than assumed by \citealt{bar08}, because the line would form on an outer and faster layer of the ejecta}. In the next paragraphs, we reflect on the implications that the asymmetries and inhomogeneities of the outer layers of the ejecta might have for the predicted spectra.

Figure~\ref{fig10} shows a map of the distribution of Fe-group elements as would be seen from an arbitrary line of sight. The column density of Fe-group elements in this direction, starting from the photosphere, is used to color the map. In order to compute the column density, the photosphere has been simply defined as the surface at which the optical depth in the line of sight is 2/3, assuming a constant opacity of 0.2~cm$^2$~g$^{-1}$ (the maximum possible column density of Fe-group elements down to the photosphere is thus $(2/3)/0.2 = 3.3$~g~cm$^{-2}$). This image shows that the distribution of Fe through the outermost layers is quite asymmetrical, hence their effect on the photon flux is expected to be latitude dependent. On average, one expects that the asymmetrical distribution of Fe should reduce the blocking of UV photons and the shifting of the flux to redder wavelengths, especially in locations where Fe is scarce such as in the upper hemisphere of this image, which should allow a better agreement between synthetic PRD spectra and observations.

The implications of the inhomogeneous chemical composition for the formation of carbon lines are illustrated in Fig.~\ref{fig11}. In this figure we have plotted a proxy of the shape of carbon lines from different views: the distribution of the absorbing mass of carbon between the photosphere and the surface as a function of the velocity projected in the line of sight (in real supernovae, the depth of the absorption lines should correlate with the mass of carbon the photons have to cross before escaping the ejecta). Obviously, the physics of line formation in supernovae is much more complicated than our simple prescription accounts for, but this is enough for the present qualitative discussion. What Fig.~\ref{fig11} shows is that the shape and peak of the carbon distribution are quite sensitive to the line of sight. The same model is able to produce broad and shallow lines and triangular-shaped deep lines depending on the line of sight. At the same time, the peak of the line can change at most by $\Delta v\sim7\,000$~km~s$^{-1}$, i.e. $\Delta\lambda\sim150${\AA}. In spite of the simplicity of these calculations, they show that the scattering in the carbon line strength due to dimensionality effects might be large and, therefore, 
conclusions obtained by comparing SNIa observations with synthetic spectra of angle-averaged versions of three-dimensional models have to be taken with care.

\section{Conclusions}

It is amazing that one-parameter spherically symmetric explosion models continue providing the best fits to SNIa observations. And it is even more amazing that elaborate three-dimensional simulations of SNIa neither reproduce one-dimensional results nor improve on them.
In spite of recent advances in 3D modeling of delayed detonations of white dwarfs, current models are still far from passing a suite of basic observational tests. The few one-dimensional spectral calculations from angle-averaged 3D SNIa models \citep[e.g.][]{bar08} did isolate several deficiencies when compared to observed SNIa spectra: the formation of undetected lines of C, an incorrect shape of the S lines, and an excessive abundance of Fe-group elements in the outer layers of the ejecta. On the other hand, observed bolometric light curves are easily reproduced by current 3D SNIa models \citep[e.g.][and the present work]{rop07c} as they depend mainly on the mass of radioactive $^{56}$Ni synthesized during the explosions. 

In this paper, we have discussed the thermal and dynamical evolution of PRD models of SNIa. The detonation phase of these models begins after accretion shock formation around a carbon-oxygen rich core, for a wide range of initial conditions. The formation of the detonation was robustly achieved in our simulations. Even though resolution limitations of our large-scale calculations preclude us to resolve the detonation wave structure, the paths followed by shocked particles in the $\rho-T$ plane surpass clearly the threshold for C-O detonation, as shown in paper~I. Thus, a confinement detonation ignition can be safely assumed in our large-scale simulations, the exception being those models in which the deflagrating phase of the explosion, previous to white dwarf pulsation, releases an amount of nuclear energy close to the binding energy of the star. 

The present PRD models score quite well on the basic observational tests quoted in Sect.~\ref{intro}:
\begin{itemize}
\item All models produced less than $\lesssim0.13$~M$_\odot$ of carbon, and only a small fraction of it was ejected with small velocities.
\item The kinetic energy and total burned mass stay in the range $1.0-1.2\times10^{51}$~erg and $1.1-1.2$~M$_\odot$, respectively \citep{maz07,woo07}. 
\item The amount of $^{56}$Ni synthesized in the explosions ranges from $0.6$~M$_\odot$ to $0.8$~M$_\odot$ that is enough to explain normal to bright events, although this range appears a little narrow to explain the differences in luminosity of the whole observational SNIa sample.
\item The mass of intermediate-mass elements is larger than $\sim0.2$~M$_\odot$ in all our models.
\item The explosions did not produce big clumps of $^{56}$Ni. Instead, radioactive as well as stable Fe-group nuclei are assembled in small bubbles pervading the outer layers of the ejecta.
\end{itemize}
The chemical profile of the PRD models lacks the degree of stratification suggested by observations. Although, in general, Si and S display an increasing mass fraction as function of the ejecta velocity, Fe-group elements represent the most abundant group except at the innermost layers made of unburnt carbon and oxygen. Stable Fe-group nuclei, synthesized at high densities during the initial deflagration phase of the explosion, are not present at the center of the ejecta as they should according to infrared observations of SNIa. This might be a weak point of the PRD scenario, as bubbles floatation and chemical composition inversion at the end of the deflagration phase are two of its basic ingredients.

We have pointed out several possible effects that the existence of asymmetries in the ejecta may have in the supernova spectra. These effects might mitigate partially some of the deficiencies found by \citet{bar08}, who calculated synthetic spectra using one-dimensional averaged versions of the present PRD models. Obviously, much work remains to be done before one can ascertain if the PRD is a reliable scenario able to account for a significant number of SNIa events.


\acknowledgments

This work has been partially supported by the MEC grants AYA2005-08013-C03, 
AYA2007-66256, by the European Union FEDER funds and by the Generalitat de 
Catalunya

\appendix

\section{Numerical techniques for SPH simulations of detonation waves}

Shocked flows present SPH simulations with specific challenges due to the presence of large pressure gradients and strongly anisotropic distributions of particles. Although there exists extensive literature devoted to such topics, we summarize in this appendix the numerical techniques we have implemented in our calculations.
The interpolating kernel used to compute the terms in the 3D SPH evolutionary equations was the cubic spline kernel $W_{ij}\left(u,h\right)$ defined in \citet{mon91}.

\subsection{Determination of the smoothing length}

We allowed for a variable number of neighbors in order to improve the behavior of the code near shock fronts, around which many particles can be crowded leading to fluctuations of the smoothing length if a fixed number of neighbors is prescribed. In our implementation, the smoothing length tracked not only the number of neighbors but also their distribution around a given particle. The smoothing length, $h_i$, of particle $i$ was defined by:
\begin{equation}
 n_0 = \sum_{j=1 \atop j\neq i}^{N}\max\left[0,1-\left(\frac{r_{ij}}{2h_i}\right)^3\right]\,,
\end{equation}

\noindent where $n_0$ is a fixed parameter, and $r_{ij}$ is the distance between particles $i$ and $j$.
This way, neighbor particles contribute to $n_0$ with a fractional number that is a decreasing function of the distance to particle $i$, while particles farther than $2h_i$ from particle $i$ do not contribute at all to the summation. When neighbor particles move across a sphere of radius $2h_i$ the summation is smoothly modified, which translates into a smooth adjustment of $h_i$ in order for the summation to match $n_0$ again. Hence, the change of $h_i$ is also a sensitive function of $\Delta t$ and fluctuations of the smoothing length are avoided. 

In \citet{gar98} we found that a minimum of $\sim40$ neighbors in 3D calculations was needed to fit the structure of hydrostatic massive white dwarfs without introducing appreciable fluctuations in the hydrostatic force. We have found that a value of order $n_0\simeq22$ results in a safe minimum of neighbors ($\gtrsim40$) that ensures an accurate determination of gradients. Working with a smaller value of $n_0$ could increase the numerical noise beyond acceptable levels, while a significantly larger number of neighbors would smooth too much the interpolated variables preventing an accurate representation of the detonation wave. In the simulations reported here we have worked with $n_0 = 21.875$.

Neighbor searching was implemented through an octal tree, the same used for the calculation of the gravity force. Unlike the above described algorithm used to compute the individual smoothing length of each particle, which applied the gather method, the evaluation of the kernel in the SPH evolutionary equations followed a symmetric momentum-conserving algorithm, combining the gather and scatter methods in the way suggested by \citet[see, e.g., their Eq.~2.15]{her89}. During the PRD simulations reported here the number of neighbors contributing to the SPH equations of any given particle lied in the range 40-90.

\subsection{Artificial viscosity}

Shock waves were handled by artificial viscosity, with the prescription of \citet{bal95} to avoid spurious entropy generation in 
regions with strong shear flows. The viscous term $\Pi_{ij}$ was computed as in \citet[][Eqs.~4.1 and 4.2]{mon92}, and later corrected by a factor $f_{ij}$ as defined in \citet[][see also \citealt{mon05}]{bal95}:
\begin{equation}
 \Pi_{ij}^\mathrm{corr} = \Pi_{ij} f_{ij} = \Pi_{ij} 0.5 \left(f_i + f_j\right)\,,
\end{equation}
\noindent where
\begin{equation}
 f_i = \frac{\left|\nabla\cdot v_i\right|}{\left|\nabla\cdot v_i\right|+\left|\nabla\times v_i\right|+0.0001c_i/h_i}
\end{equation}
\noindent with $\nabla\cdot v_i$ and $\nabla\times v_i$ the divergence and the curl of the velocity field, which we compute following \citet[][Eqs.~2.10 and 2.12]{mon92}, and $c_i$ the sound velocity of particle $i$.

\subsection{Anisotropic interpolation kernel}

Although the use of artificial viscosity is 
not optimal in comparison with Riemann solvers, the use of anisotropic 
interpolation kernels greatly improves the ability of SPH codes to deal with 
shock waves. We have adapted a 3D ellipsoidal kernel \citep{sha96,owe98} to our problem. When a 
detonation front is detected, the (otherwise spherically symmetric) interpolating kernel takes the shape of two half oblate spheroids joined at the equator (see Fig.~\ref{figkernel}), sharing an axis of rotational symmetry that coincides with the direction of the local pressure gradient (computed with the standard spherically symmetric kernel). The size of each semi-minor axis, $2h_+$ ($2h_-$), is determined in three steps. A first value is obtained through:
\begin{equation}
 h_{\pm} = \max\left( \delta_1\frac{P_i}{\left|\nabla P_{\pm}\right|}, \delta_2\frac{10^{10}~\mathrm{K}}{\left|\nabla T_{\pm}\right|} \right)\,, \label{eqa5}
\end{equation}
\noindent where $\nabla P_+$ and $\nabla T_+$ ($\nabla P_-$ and $\nabla T_-$) are the pressure and temperature gradients at the position of particle $i$ computed using only particles located in the forward (backward) direction of $\nabla P_i$, and $\delta_1$ and $\delta_2$ are two coefficients of order unity that we take here as $\delta_1=\delta_2=0.13$. The aim of the differentiation between both directions of $\nabla P_i$ is that the presence of a discontinuity in front (back) of a particle does not limit its capability to contribute to the determination of the smoothed quantities in the backward (forward) direction. 
In a second step we require that the flow be sonic by recomputing $h_{\pm}$ through:
\begin{equation}
 h_{\pm} = h_{\pm}\max\left(1,0.01\frac{c_i}{v_i}\right)\,,
\end{equation}
\noindent where $c_i$ and $v_i$ are the sound velocity and the velocity of particle $i$, respectively. The last equation together with the pressure term in Eq.~\ref{eqa5} discard the activation of the anisotropic kernel in deflagration waves.
Finally, the value of $h_{\pm}$ is constrained to be in the range $h\geq h_{\pm}> 0.1h$. 
In regions where there are no shocks the kernel reduces to the standard spherically symmetric kernel. 
In this way the spatial resolution at the detonation front improved, on average, by a factor $\sim2-3$. 

To compute the anisotropic kernel $W_{ij}^{\mathrm{ani}}\left(u\right)$ at the position of a neighbor particle $j$ (see Fig.~\ref{figkernel}) we use the cubic spline kernel \citep{mon91} as a function of the ratio $u$ between the particles distance $r_{ij}$ and the size of the kernel in the direction of $\vec{r}_{ij}$: $u=2r_{ij}/d$, with:
\begin{equation}
 d = 2\frac{hh_{\pm}}{\sqrt{h_{\pm}^2+\left(h^2-h_{\pm}^2\right)\cos^2 a}}\,,
\end{equation}
\noindent and $a$ the angle between $\vec{r}_{ij}$ and $\nabla P$. The anisotropic kernel is finally normalized by a factor 
\begin{equation}
 \frac{2\pi}{h^2\left(h_+ + h_-\right)}\,,
\end{equation}
\noindent so that 
\begin{equation}
 \int_{0}^{\infty} W_{ij}^{\mathrm{ani}}\,\mathrm{d}\vec{r} = 1\,.
\end{equation}

\begin{figure}
\plotone{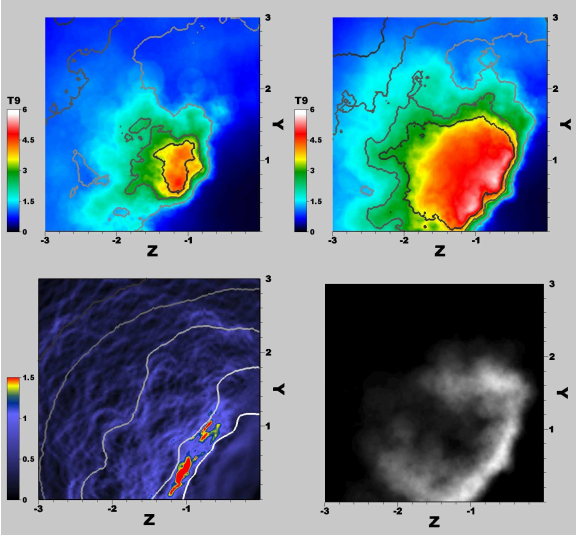}
\caption{Detonation ignition in model PRD14. Temperature maps and fuel mass fraction contours are shown in a slice around the location of detonation formation: at $t=4.0$~s, shortly after ignition (top left image), and 0.1~s later (top right image). The bottom left image is a reconstruction of the hydrostatic force ($\frac{\nabla P}{\rho}$, color map in units of $10^{10}$~cm~s$^{-2}$) with density contours ($\mathrm{log}\rho (\mathrm{g~cm}^{-3})= 8$, 7.5, 7, and 6.5 from the center outwards) at $t=4.1$~s. The bottom right image shows in white the region in which a detonation wave has been detected by the detonation capture algorithm at $t=4.1$~s.
Axes units, temperature scale and fuel mass fractions represented are the same as in Fig.~\ref{newfig3}. The slice is a zoom on the $yz$ plane of the octants shown in Figs.~\ref{newfig3} and \ref{figtem}
}\label{newfig4}
\end{figure}

\begin{figure}
\plotone{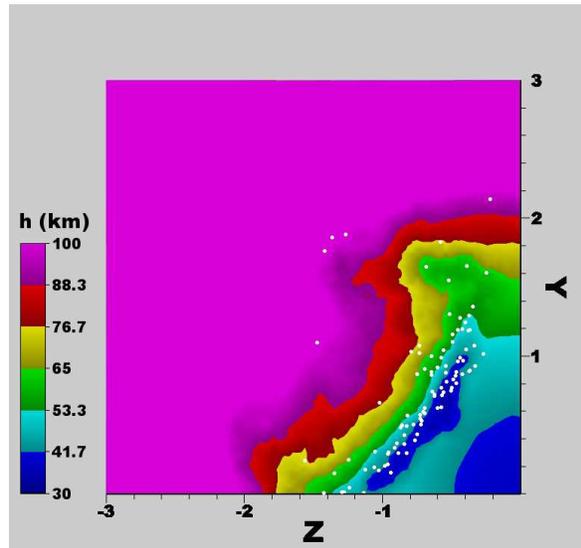}
\caption{Maximum resolution of the SPH calculation at $t=4.1$~s in the same slice shown in Fig.~\ref{newfig4}. The color map represents the smoothing length, $h$, in km. The white points give the actual position of the SPH particles for which the detonation capture algorithm has been activated and the anisotropic smoothing kernel is at work with its semiminor axis aligned with the pressure gradient
}\label{newfig5}
\end{figure}

\begin{figure}
\plotone{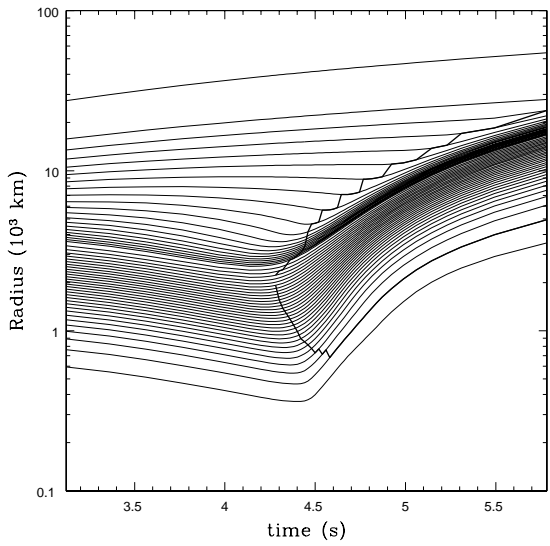}
\caption{Dynamical evolution of model PRD14 shortly before and during detonation propagation. Each thin line shows the radius of a spherical layer enclosing a constant mass. The thick lines show the inner and outer boundaries of the detonated layers, that are defined for the purposes of the present figure as those layers in which the mass of fuel processed by the detonation wave is larger than the mass of remaining fuel
}\label{newfig2}
\end{figure}

\begin{figure}
\plotone{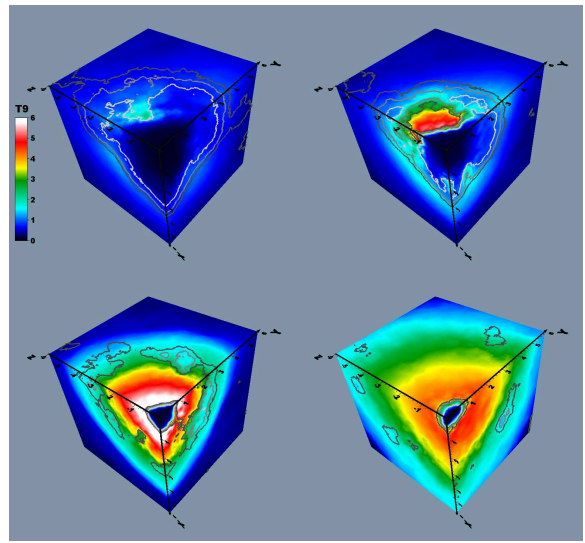}
\caption{Evolution of temperature (color map, in units of $10^9$~K) and fuel mass fraction (contours) in an octant of model PRD14 
during the detonation phase. The C+O mass fraction contours shown are $X=0.3$ (black), 0.5 (dark gray), and 0.8 (light gray). Axes units are thousands of km. Times shown are (from left to right and top to bottom): $t=3.8$, 4.1, 4.4, and 4.7~s
}\label{figtem}
\end{figure}

\begin{figure}
\plotone{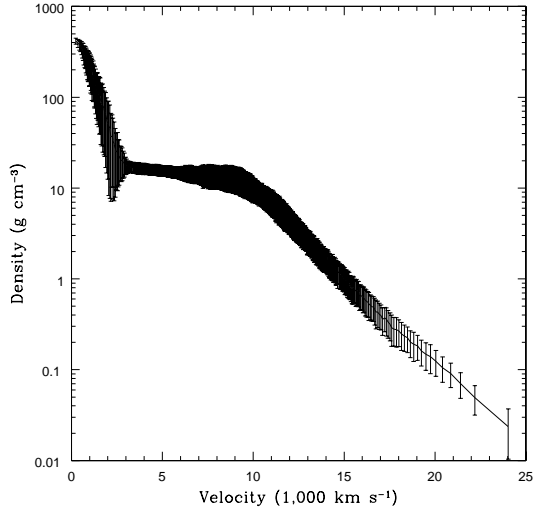}
\caption{Final angle-averaged density profile as a function of radius for model PRD14. The 
distribution of points resulting from the SPH calculation has been divided 
into 1\,000 radial shells. The mean density and the dispersion within each 
shell are shown
}\label{figdens}
\end{figure}

\begin{figure}
\plotone{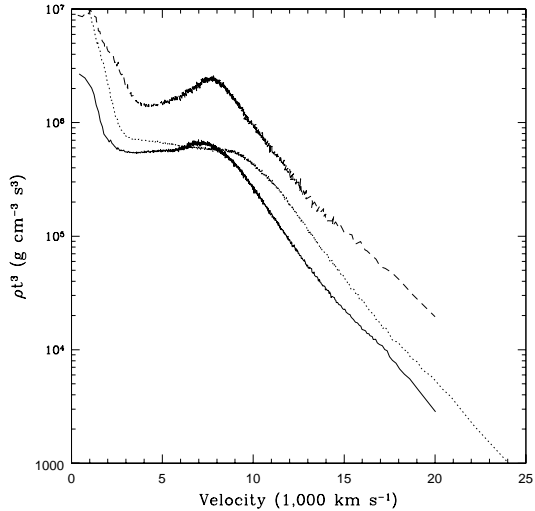}
\caption{Final angle-averaged density profiles of models PRD14 at $t=34$~s (dotted line), PRD18 at 1.5~h (solid line), and PRD26 at 20~s (dashed line). The density is shown here factorized by $t^3$ in order to make the densities comparable
}\label{figdens2}
\end{figure}

\begin{figure}
\plotone{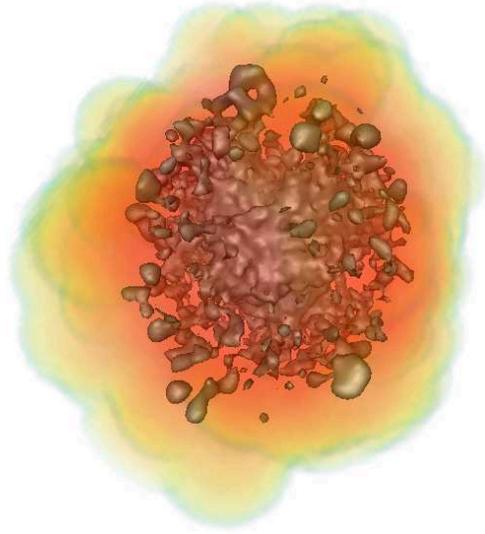}
\caption{Rendering of the final distribution of $^{56}$Ni in model
PRD14. The isosurface defined by  $X(^{56}\mathrm{Ni})=0.5$ together with a volume rendering of the ejecta density are shown
}\label{fig5}
\end{figure}

\begin{figure}
\plotone{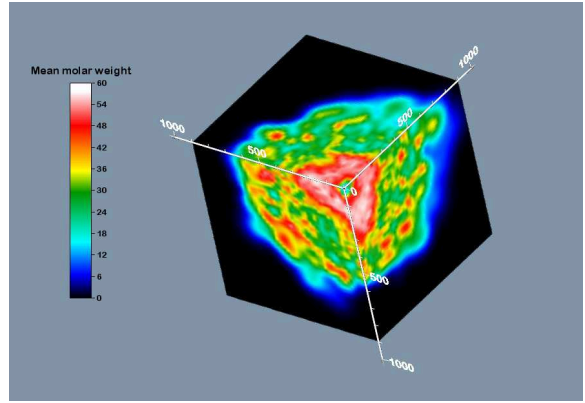}
\caption{Final chemical composition in an octant of model PRD14. It is
interesting to note that 
the chemical structure at the end of the pulsating phase (Fig.~3 in Paper~I) 
left no imprint on the final distribution of elements. The detonation led 
to a quite symmetric distribution with some degree of chemical 
stratification, with C-O confined to the outskirts of the ejecta, Si-S dominant
at intermediate distances, and Fe-group nuclei concentrated in the inner ejecta
but also present in clumps distributed through the outer ejecta. The axis units are
thousands of km
}\label{fig05}
\end{figure}

\begin{figure}
\plotone{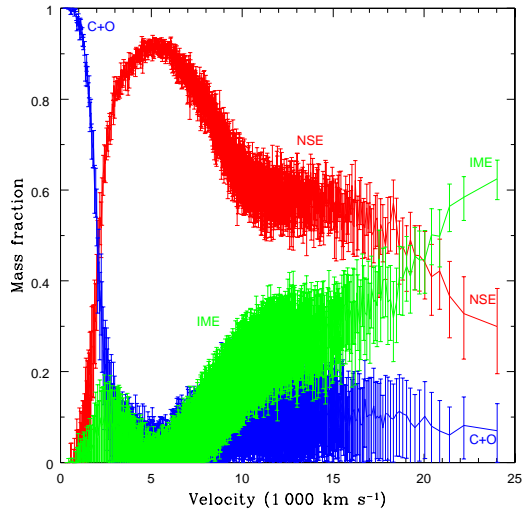}
\caption{Final angle-averaged chemical profiles of model PRD14 as a function of the velocity. The 
distribution of points resulting from the SPH calculation has been divided 
into 1\,000 radial shells. The mean mass fraction and the dispersion within each 
shell are shown. The fuel (C+O) mass fraction is painted in blue, that of intermediate-mass elements in green, and that of Fe-group elements in red
}\label{figcomp1}
\end{figure}

\begin{figure}
\plotone{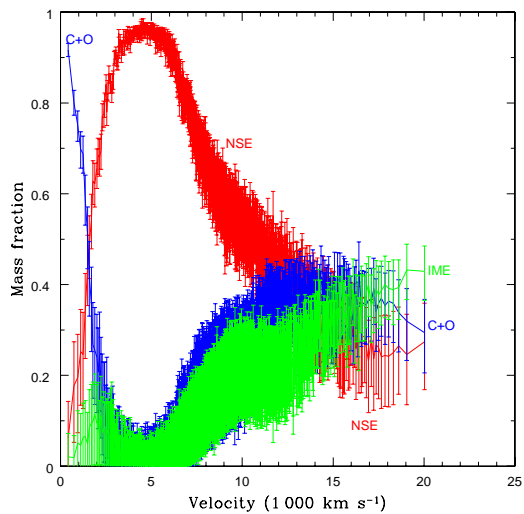}
\caption{Same as Fig.~\ref{figcomp1}, but for model PRD18
}\label{figcomp2}
\end{figure}

\begin{figure}
\plotone{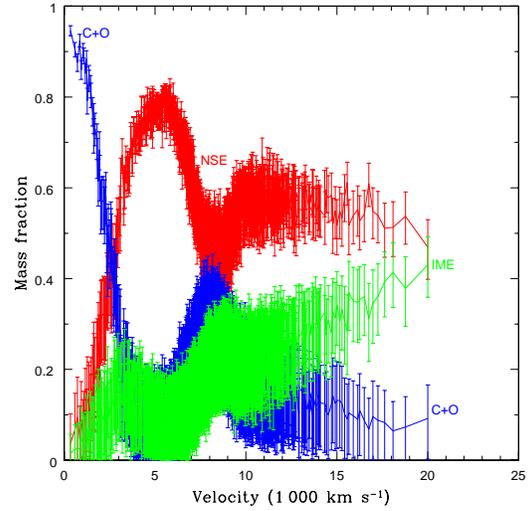}
\caption{Same as Fig.~\ref{figcomp1}, but for model PRD26
}\label{figcomp3}
\end{figure}

\begin{figure}
\plotone{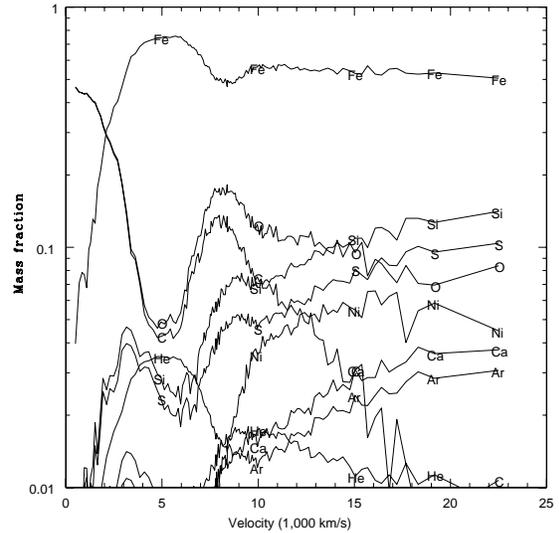}
\caption{Abundances at $t=\infty$ in the ejecta of model PRD14 as a function of
the velocity. The abundances are angular averaged in 130 spherical shells
}\label{fig12}
\end{figure}

\begin{figure}
\plotone{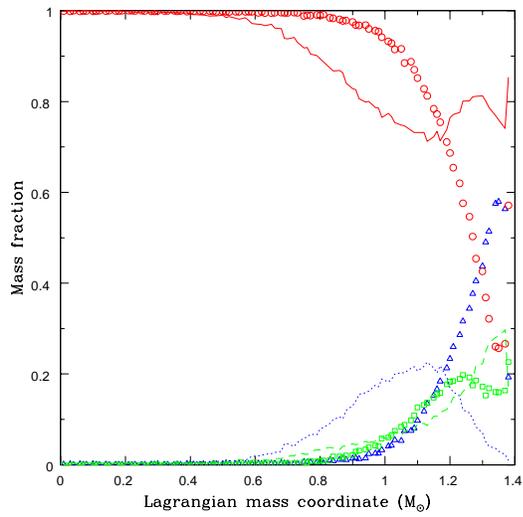}
\caption{Chemical profile of the delayed detonation model TURB7 just after the end of the detonation phase (points) and in the homologous expansion phase (lines: $t=62$~s). Carbon and oxygen are represented by blue triangles and the dotted line, intermediate-mass elements by green squares and the dashed line, and Fe-peak elements by red circles and the solid line
}\label{vturb}
\end{figure}

\begin{figure}
\plotone{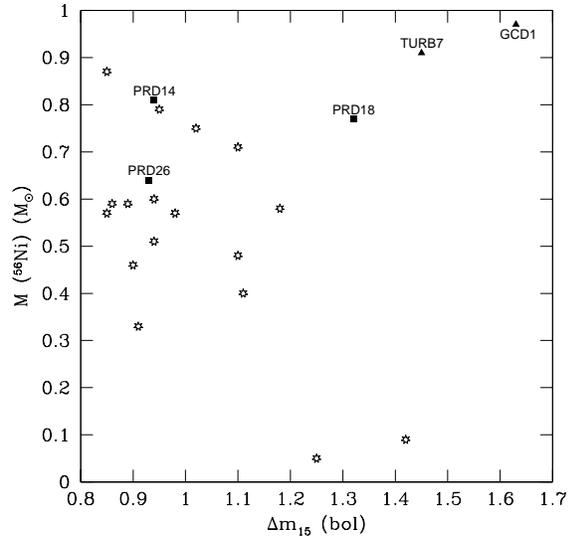}
\caption{Mass of $^{56}$Ni synthesized in SNIa explosions vs decrease in bolometric magnitude 15 days after maximum. Starred symbols are data for a sample of well observed SNIa from \citet{str06}, full squares belong to the Pulsating Reverse Detonation models, while full triangles represent models GCD1 and TURB7
}\label{figm15m56}
\end{figure}

\begin{figure}
\plotone{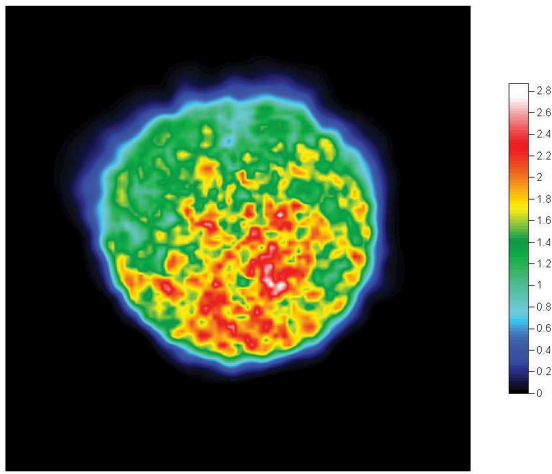}
\caption{Column density of Fe-group elements for model PRD14 at day 25 after
thermal runaway. The column density (in g$\cdot$cm$^{-2}$) of Fe-group elements
between the photosphere and the observer is color-coded for an arbitrary line of sight. White and red colors mark places for which the material above the photosphere is nearly pure Fe, while blue and green indicate the contrary: that Fe is scarce. 
}\label{fig10}
\end{figure}

\begin{figure}
\plotone{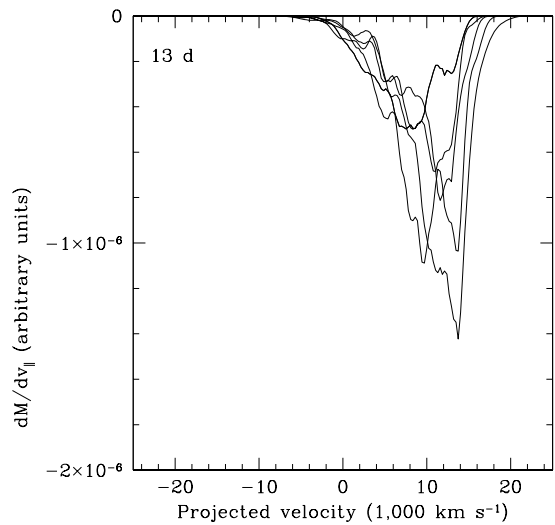}
\caption{Distribution of the mass of carbon between the photosphere and the observer as a function
of the velocity in the line of sight for model PRD14 at day 13 after
thermal runaway. The different curves belong to 5 arbitrary lines of sight. For this calculation the ejecta has been mapped to a three-dimensional grid in spherical coordinates, and the location of the photosphere has been estimated as for Fig.~\ref{fig10} (see text)
}\label{fig11}
\end{figure}

\begin{figure}
\plotone{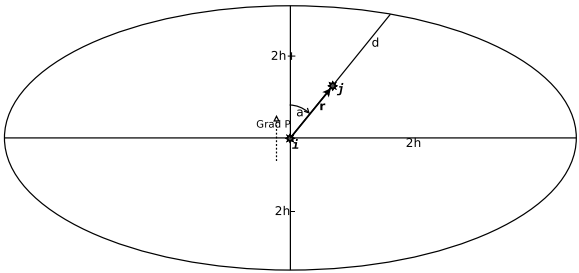}
\caption{Geometry of the ellipsoidal kernel of particle $i$. The semi-major axis along a plane normal to the pressure gradient is the smoothing length $h$, while the semi-minor axis is aligned with $\nabla p$. In our implementation we use two different values of the semi-minor axis, one in each direction of the pressure gradient: $h_+$ and $h_-$ (see text for details). The kernel at the position of a neighbor particle $j$, depends on the angle, $a$, between the relative position vector $\vec{r}$ and the direction of the semi-minor axis, and on the ratio $u$ of the particles distance to the size of the ellipsoid, $d$, in the direction of $\vec{r}$: $u=2r/d$
}\label{figkernel}
\end{figure}

The ability of our device to follow the propagation of detonation waves was demonstrated in \citet[Fig. 1]{gar99}, where more details of the numerical method are given, and in Figs.~\ref{newfig4} and \ref{newfig5} in the present paper. A problem with ellipsoidal kernels is that they do not guarantee conservation of energy and angular momentum. In fact, during the detonation phase the violation of energy conservation reached $\sim5\%$ of the final kinetic energy, while the net angular momentum at the end of the simulation was negligibly small \citep[for reference figures see Table~1 in][]{bra08}.

\bibliographystyle{aa}
\bibliography{../../../ebg}

\clearpage
\centering
\begin{deluxetable}{lcccccc}
\tabletypesize{\scriptsize}
\tablecaption{SPH 3D delayed detonation models compared with other works}
\tablecolumns{7}
\tablewidth{0pt}
\tablehead{
\colhead{Model} &
\colhead{$M_\mathrm{defl}$} &
\colhead{$K$} &
\colhead{$M \left(^{56}\mathrm{Ni}\right)$} &
\colhead{$M \left(\mathrm{NSE}\right)$} &
\colhead{$M \left(\mathrm{C+O}\right)$} &
\colhead{$M \left(\mathrm{IME}\right)$} \\
 & (M$_\odot$) & ($10^{51}$~ergs) & (M$_\odot$) & (M$_\odot$) & (M$_\odot$) & (M$_\odot$) \\
}
\startdata
GCD1 & 0.06 & 1.51 & 0.97 & 1.21 & 0.04 & 0.10 \\
TURB7 & 0.40 & 1.49 & 0.91 & 1.19 & 0.09 & 0.07 \\
Y100\tablenotemark{a} & & 1.52 & $<1.19$ & 1.19 & 0.04 & 0.13 \\
DD\_005\tablenotemark{b} & 0.38 & 1.52 & $<1.14$ & 1.14 & 0.04 & 0.22 \\
\enddata
\tablenotetext{a}{\citet{pl07}}
\tablenotetext{b}{\citet{rop07}}
\label{tab1}
\end{deluxetable}

\clearpage
\centering
\begin{deluxetable}{lcccccccc}
\tabletypesize{\scriptsize}
\tablecaption{Properties of the PRD models
}
\tablecolumns{9}
\tablewidth{0pt}
\tablehead{
\colhead{Model} &
\colhead{$M_\mathrm{defl}$} &
\colhead{$M_\mathrm{acc}$\tablenotemark{a}} &
\colhead{$v_\mathrm{acc}$\tablenotemark{b}} &
\colhead{$v_\mathrm{deto}$\tablenotemark{c}} &
\colhead{$K$} &
\colhead{$M \left(^{56}\mathrm{Ni}\right)$} &
\colhead{$L_\mathrm{max}$\tablenotemark{d}} &
\colhead{$\Delta m_{15}$\tablenotemark{e}} \\
 & (M$_\odot$) & (M$_\odot$) & ($1\,000$~km~s$^{-1}$) & ($1\,000$~km~s$^{-1}$) & ($10^{51}$~erg) & (M$_\odot$) & (erg~s$^{-1}$) & \\
}
\startdata
PRD14 & 0.14 & 1.26 & 14.0 & 9.8 & 1.22 & 0.81 & $2.4\times10^{43}$ & 0.94 \\
PRD18\tablenotemark{f} & 0.18 & 1.15 & 11.0 & 9.3 & 1.05 & 0.77 & $2.0\times10^{43}$ & 1.32 \\
PRD26 & 0.26 & 1.05 & 10.0 & 8.0 & 1.01 & 0.64 & $1.9\times10^{43}$ & 0.93 \\
PRD29 & 0.29 & 0.83 & - & - & - & - & - & - \\
\enddata
\tablenotetext{a}{ {Lagrangian} mass of the accretion shock}
\tablenotetext{b}{Final ejecta velocity at the Lagrangian mass $M_\mathrm{acc}$}
\tablenotetext{c}{Final ejecta velocity at the Lagrangian mass at which a confined detonation was ignited}
\tablenotetext{d}{Maximum luminosity}
\tablenotetext{e}{Decline in bolometric magnitude 15 days after bolometric maximum}
\tablenotetext{f}{The nucleosynthesis of model PRD18, including the $^{56}$Ni
yield, has been computed with a post-processing code and hence differs slightly
from the figures given by the alpha-network built into the SPH code, which were
provided in \cite{bra06}}
\label{tab2}
\end{deluxetable}

\clearpage
\centering
\begin{deluxetable}{lccc}
\tabletypesize{\scriptsize}
\tablecaption{Chemical composition of the ejecta in solar masses}
\tablecolumns{4}
\tablewidth{0pt}
\tablehead{
\colhead{Model} &
\colhead{PRD14} &
\colhead{PRD18} &
\colhead{PRD26} \\
}
\startdata
\element{}{}{C}  & 0.061  & 0.101  & 0.122  \\ 
\element{}{}{O}  & 0.105  & 0.149  & 0.169  \\ 
\element{}{}{Ne} & 0.010  & 0.015  & 0.015  \\
\element{}{}{Mg} & 0.025  & 0.027  & 0.028  \\
\element{}{}{Si} & 0.118  & 0.102  & 0.138  \\
\element{}{}{S}  & 0.055  & 0.052  & 0.068  \\
\element{}{}{Ar} & 0.011  & 0.011  & 0.014  \\
\element{}{}{Ca} & 0.010  & 0.011  & 0.013  \\
\element{}{}{Cr} & 0.005  & 0.007  & 0.006  \\
\element{}{}{Mn} & 0.001  & 0.002  & 0.001  \\
\element{}{}{Fe} & 0.878  & 0.815  & 0.720  \\
\element{}{}{Ni} & 0.054  & 0.046  & 0.051  \\
\enddata
\label{tab3}
\end{deluxetable}

\clearpage



\clearpage

\end{document}